\documentclass[3p,times]{elsarticle}

\usepackage{color}

\usepackage{ecrc}
\usepackage{amsmath}

\volume{00}

\firstpage{1}

\journalname{Physics Procedia}

\runauth{}

\jid{procs}

\jnltitlelogo{Physics Procedia}

\CopyrightLine{2011}{Published by Elsevier Ltd.}

\usepackage{amssymb}

\usepackage[figuresright]{rotating}

\begin{document}

\begin{frontmatter}



\dochead{}

\title{Model-independent Analyses of Dark-Matter Particle Interactions}


\author{Nikhil Anand$^{a,b}$}
\author{A. Liam Fitzpatrick$^{c}$}
\author{W. C. Haxton$^{a,d}$}

\address[a]{Department of Physics, University of California, Berkeley, CA 94720}
\address[b]{Department of Physics, Johns Hopkins University, Laurel, MD 20723}
\address[c]{Stanford Institute for Theoretical Physics, Stanford University, Stanford, CA 94305}
\address[d]{Lawrence Berkeley National Laboratory, Berkeley, CA 94720}

\begin{abstract}
A model-independent treatment of dark-matter particle elastic scattering has been developed,
yielding the most general interaction for WIMP-nucleon low-energy scattering,
and the resulting amplitude has been embedded into the nucleus, taking into account
the selection rules imposed by parity and time-reversal.  One finds that, in contrast to the
usual spin-independent/spin-dependent  (SI/SD) formulation, the resulting cross
section contains six independent nuclear response functions, three of which are associated
with possible velocity-dependent interactions. We find that current
experiments are four orders of magnitude more sensitive to derivative couplings
than is apparent in the standard SI/SD treatment, which necessarily associated such interactions with cross sections
proportional to $v_T^2 \sim 10^{-6}$, where $v_T$ is the WIMP velocity relative to the center of mass 
of the nuclear target.
\end{abstract}


\end{frontmatter}


\section{Introduction}
\label{sec:intro}
The experimental community is engaged in an effort to characterize the properties of
dark matter using nuclear recoil as the signal for the elastic scattering of heavy dark
matter particles off nuclear targets.  While astrophysics has established certain basic properties of
dark matter -- that it is long-lived or stable, cold or warm, gravitationally active, but without
strong couplings to itself or to baryons -- the nature of the interactions of
dark-matter particles with ordinary matter is otherwise unknown \cite{Freese2013,Bertone2005}.  Among the leading
dark-matter candidates are weakly interacting massive particles (WIMPs): it was
noted long ago that such weak-scale particles could be produced in the  Big Bang 
with abundances on the order of that required by observation, $\Omega_{DM} \sim 0.27$.
Additional motivation for WIMPs comes from expectations that new physics at the
mass-generation scale of the standard model might naturally lead to 
new particles of mass $\sim$ 10 GeV - 10 TeV.

One of the promising methods for detecting WIMPs is via their elastic
scattering with heavy nuclei (``direct detection") \cite{Goodman1985,Drukier1986,Jungman1996}.
The WIMPs would be slowly moving today, with velocities $v_T \sim 10^{-3}$,
and consequently with momenta $q \sim 100$ MeV/c  and kinetic energies 
$\epsilon \sim$ 50 keV.
The momentum scale is on the order of the inverse size of the nucleus,
$\hbar /R_\mathrm{nucleus} \sim \hbar /(1.2 fm~ A^{1/3}) \sim 30$ MeV/c, while $\epsilon$
is small compared to the typical excitation energies in a nucleus of $\sim$ 1 MeV.  
Consequently
\begin{itemize}
\item  the detection of WIMP elastic interactions with a nucleus 
requires sensitivity to recoil energies on the order of 10s of keV; 
\item  inelastic interactions need only be considered in unusual cases where a target
nucleus has an excited state within $\sim$ 100 keV of the ground state; and 
\item a proper quantum mechanical  treatment of the elastic scattering cross section
should take into account the size of the nucleus, as $q R_\mathrm{nucleus} \gtrsim 1$.
\end{itemize}
Because the WIMP will, in most cases, only scatter elastically, one also sees that
parity and time-reversal selection rules that operate for diagonal matrix elements
will limit what can be learned in direct detection experiments.

While we know little about dark matter interactions with ordinary matter,
their possible associations with electroweak interactions suggests using the standard model as a guide.
In electromagnetism, elastic scattering can
occur through charge or magnetic interactions.  Both interactions involve nontrivial isospin --
the charge coupling is only to protons, while the magnetic coupling involves the distinct proton
and neutron magnetic moments. Magnetic elastic scattering occurs through two interfering 
three-vector operators, spin $\vec{\sigma}(i)$ and orbital angular momentum $\vec{\ell}(i)$.
For weak interactions, the weak charge operator couples primarily to neutrons, while the axial-charge 
operator $\vec{\sigma}(i) \cdot \vec{p}(i)$ makes effectively no contribution to elastic scattering, apart from small recoil corrections,
due to the constraints imposed by parity and time-reversal invariance.  One might expect,
consequently, that the WIMP-nuclear interaction will involve a variety of operators
as well as couplings that depend on isospin.

In part for historical reasons, WIMP elastic scattering experiments are most often
analyzed by assuming the interaction is simpler than those described above: isoscalar, 
coupled either to the nucleon number operator $1(i)$ (spin-independent or SI)  or the 
nucleon spin $\sigma(i)$ (spin-dependent or SD) \cite{Goodman1985,Smith1990,Lewin1996}.
These are the operators for a point nucleus.  While a form factor is often introduced to account 
phenomenologically for the fact that the momentum transfer is large on the nuclear scale, 
the quantum mechanical consequences  of $o(1)$ operators like $\vec{q} \cdot \vec{r}(i)$
have been largely neglected.

Recently there have been efforts to treat the WIMP-nucleon interaction in more generality,
using the tools of effective field theory (EFT) \cite{Fan2010,Liam2013,Hill2013,Nikhil2014}.  We 
describe the approach of \cite{Liam2013,Nikhil2014} in Sec. \ref{sec:response} and
its consequences for WIMP-nucleus elastic scattering.  Consistent with general symmetry
arguments, six independent nuclear response functions are identified, in contrast to the 
two assumed in SI/SD treatments.  The new operators are associated with derivative couplings,
where a proper treatment of $\vec{q} \cdot \vec{r}(i)$ is essential due to the need to identify
associated parity- and time-reversal-conserving elastic operators.  When this is done, we find
that velocity-dependent interactions lead to cross sections
$\sim q^2/m_N^2 ~G_F^2\sim 10^{-2} ~G_F^2$, where $m_N$
in the nucleon mass and $G_F$ the weak coupling constant, in contrast to the SI/SD result, 
$\sim v_T^2~ G_F^2 \sim 10^{-6} ~G_F^2$.  Our effective theory treatment 
shows that much more can be learned about the properties of WIMP dark matter from
elastic scattering experiments than is generally appreciated.  However, it also shows that a 
greater variety of experiments will be necessary to extract this information and to 
eliminate possible sources of confusion, when competing experiments are compared.\\

\section{The Nuclear Elastic Response from Effective Theory}
\label{sec:response}
Here we summarize the effective theory construction of the WIMP-nucleon interaction of Ref. \cite{Liam2013,Nikhil2014}.
Details can be found in the original papers. 
The Lagrangian density for the scattering of a WIMP off a nucleon is taken to have the form
\begin{equation}
\mathcal{L}_{\textrm{int}}(\vec{x}) =c ~ \Psi^*_\chi(\vec{x}) {{\mathcal{O}}}_\chi \Psi_\chi(\vec{x})~ \Psi^*_N(\vec{x}) {{\mathcal{O}}}_N \Psi_N(\vec{x}),
\end{equation}
where the $\Psi(\vec{x})$ are nonrelativistic fields and where the WIMP and
nucleon operators  ${{\mathcal{O}}}_\chi$ and $ {{\mathcal{O}}}_N$ may have vector indices.
The operators ${{\mathcal{O}}}_\chi$ and ${{\mathcal{O}}}_N$ are then allowed to take on their most 
general form, constrained by imposing relevant symmetries. 
The construction was done in the nonrelativistic limit to second order in the momenta.
Thus the relevant operators are those appropriate for use with
Pauli spinors.  The Galilean-invariant amplitudes take the form
\begin{equation}
 \sum_{i=1}^{\cal{N}} \left( c_i^\mathrm{n} ~{{\mathcal{O}}}_i^{\mathrm{\,n}} + c_i^{\mathrm{p}}~{{\mathcal{O}}}_i^{\mathrm{\,p}} \right),
\label{eq:Lag}
\end{equation}
where the coupling coefficients $c_i$ may be different for proton and neutrons.
The number $\cal{N}$ of such operators $\mathcal{O}_i$ -- which have the product form $\mathcal{O}^i_\chi \otimes\mathcal{O}^i_N$ --  depends 
on the generality of the particle physics description.  

The WIMP and nucleon operators for this construction include $1_\chi$ and
$1_N$,  the three-vectors $\vec{S}_\chi$ and $\vec{S}_N$, and the momenta of the WIMP and nucleon. Of the four momenta involved in the scattering (two incoming and two outgoing), only two combinations are physically relevant due to inertial frame-independence and momentum conservation.  It is convenient to work with the frame-invariant quantities, the momentum transfer $\vec{q}$ and the WIMP-nucleon
relative velocity,
\begin{equation}
\vec{v} \equiv \vec{v}_{\chi, \rm in} - \vec{v}_{N, \rm in}.
\end{equation}
The Hermitian form of this velocity is
\begin{equation}
\vec{v}^\perp = \vec{v} + \frac{\vec{q}}{2\mu_N} = {1 \over 2} \left( \vec{v}_{\chi,\mathrm{in}}+\vec{v}_{\chi,\mathrm{out}} - \vec{v}_{N,\mathrm{in}} - \vec{v}_{N,\mathrm{out}} \right) =
 {1 \over 2} \left( {\vec{p} \over m_\chi}+{\vec{p}^{\, \prime} \over m_\chi} - {\vec{k} \over m_N} - {\vec{k}^{\, \prime} \over m_N} \right)
\label{eq:vperp}
\end{equation}
which satisfies $\vec{v}^\perp \cdot \vec{q} =0$ as a consequence of energy conservation.  Here $\mu_N$ is the WIMP-nucleon reduced mass, $\vec{p}$ ($\vec{p}^{\,\prime}$)
is the incoming (outgoing) WIMP three-momentum, and $\vec{k}$ ($\vec{k}^\prime$) the incoming
(outgoing) nucleon three-momentum.
It was shown in \cite{Liam2013} that operators are guaranteed to be Hermitian if they are built out of the following four  three-vectors,
 \begin{equation}
 \vec{S}_\chi~~~~~~~ \vec{S}_N~~~~~~~\vec{v}^\perp~~~~~~~i {\vec{q} \over m_N}
 \end{equation}
The nucleon mass $m_N$ has been introduced as a convenient scale to render $\vec{q}/m_N$ and the constructed
${{\mathcal{O}}}_i$ dimensionless: the choice of this scale is not arbitrary, but is instead connected with the velocity
operator when that operator is embedded in a nucleus, as we discuss later.  One finds to second order in velocities and
momenta,
\begin{eqnarray}
{\mathcal{O}}_1 &=& 1_\chi 1_N \nonumber \\
{\mathcal{O}}_2 &=& (v^\perp)^2 \nonumber \\
 {\mathcal{O}}_3 &=& i \vec{S}_N \cdot ({\vec{q} \over m_N} \times \vec{v}^\perp) \nonumber \\
{\mathcal{O}}_4 &=& \vec{S}_\chi \cdot \vec{S}_N \nonumber \\
{\mathcal{O}}_5 &=& i \vec{S}_\chi \cdot ({\vec{q} \over m_N} \times \vec{v}^\perp)  \nonumber \\
 {\mathcal{O}}_6&=& (\vec{S}_\chi \cdot {\vec{q} \over m_N}) (\vec{S}_N \cdot {\vec{q} \over m_N}) \nonumber  \\
{\mathcal{O}}_7 &=& \vec{S}_N \cdot \vec{v}^\perp \nonumber \\
{\mathcal{O}}_8 &=& \vec{S}_\chi \cdot \vec{v}^\perp \nonumber \\ 
 {\mathcal{O}}_9 &=& i \vec{S}_\chi \cdot (\vec{S}_N \times {\vec{q} \over m_N}) \nonumber \\
{\mathcal{O}}_{10} &=& i \vec{S}_N \cdot {\vec{q} \over m_N}  \nonumber \\
{\mathcal{O}}_{11} &=& i \vec{S}_\chi \cdot {\vec{q} \over m_N} \nonumber \\
{\mathcal{O}}_{12} &=& \vec{S}_\chi \cdot (\vec{S}_N \times \vec{v}^\perp) \nonumber \\ 
{\mathcal{O}}_{13} &=&i (\vec{S}_\chi \cdot \vec{v}^\perp  ) (  \vec{S}_N \cdot {\vec{q} \over m_N})  \nonumber  \\
{\mathcal{O}}_{14} &=& i ( \vec{S}_\chi \cdot {\vec{q} \over m_N})(  \vec{S}_N \cdot \vec{v}^\perp )  \nonumber \\
{\mathcal{O}}_{15} &=& - ( \vec{S}_\chi \cdot {\vec{q} \over m_N}) ((\vec{S}_N \times \vec{v}^\perp) \cdot {\vec{q} \over m_N}) 
\label{eq:ops}
\end{eqnarray}
The first eleven of these operators can be generated by interactions
involving only spin-0 or spin-1 mediators, and were 
discussed in \cite{Liam2013}.  One of these,
${\mathcal{O}}_2$, was later eliminated from consideration in \cite{Nikhil2014}, as it cannot be 
obtained from the leading-order non-relativistic reduction
of a manifestly relativistic operator.  The remaining operators are associated with more
complicated exchanges.  These operators can be labeled
as  LO, NLO, and N$^2$LO, according to the total number of
momenta and velocities they contain,
corresponding to total cross sections that scale as $v_T^0$, $v_T^2$, or $v_T^4$, where $v_T$ is the WIMP
velocity in the laboratory frame (though, below, we stress that this scaling is accompanied by
additional factors that arise from the treatment of velocity-dependent interactions).  Operator ${\mathcal{O}}_{15}$ is
cubic in velocities and momenta, generating a total cross section of order $v_T^6$ (N$^3$LO).
It is retained because it arises as the leading-order nonrelativistic limit of a convariant
four-fermion interaction, as discussed in \cite{Nikhil2014}.
Each operator
can have distinct couplings to protons and neutrons.   Introducing isospin, which
is also useful as an approximate symmetry of the nuclear wave functions, the general form
of the effective theory interaction becomes
\begin{equation}
 \sum_{i=1}^{15} ( c_i^{0} 1 + c_i^{1} \tau_3) {\mathcal{O}}_i = \sum_{\tau=0,1} \sum_{i=1}^{15} c_i^\tau {\mathcal{O}}_i t^\tau,
~~~ c_2^\tau \equiv 0, \nonumber \\
\label{eq:Lagrangian}
\end{equation}
where $c_i^{0} = {1 \over 2} (c_i^{\mathrm{p}}+c_i^{\mathrm{n}})$, $c_i^{1} = {1 \over 2} (c_i^{\mathrm{p}}-c_i^{\mathrm{n}})$, $t^0 = 1$, and
$t^1=\tau_3$.
Thus the EFT has a total of 28 parameters, associated with 14 space/spin operators each
of which can have distinct couplings to protons and neutrons. 

This interaction can then be embedded in the nucleus.  The procedure followed in
\cite{Liam2013,Nikhil2014} assumes that the nuclear interaction is the 
sum of the WIMP interactions with the individual nucleons in the nucleus.  The nuclear
operators then involve a convolution of the ${\mathcal{O}}_i$, whose momenta must now be treated
as local operators appropriate for bound nucleons, with the plane wave associated with
the WIMP scattering, which is an angular and radial operator that can be decomposed 
with standard spherical harmonic methods.  Because momentum transfers are typically
comparable to the inverse nuclear size, it is crucial to carry through such a multipole
decomposition in order to identify the nuclear responses associated with the various
$c_i$s.   The scattering probability is given by the square of the (Galilean) invariant
amplitude $\mathcal{M}$, a product of WIMP and nuclear matrix elements, averaged over initial
WIMP and nuclear magnetic quantum numbers $M_\chi$ and $M_N$, and summed over final 
magnetic quantum numbers.  

The result can be organized
in a way that factorizes the particle and nuclear physics
 \begin{equation}
\frac{1}{2j_\chi + 1}\frac{1}{2j_N+1}\sum_{\textrm{spins}} |\mathcal{M}|^2 \equiv \sum_{k} 
\sum_{\tau,\tau^\prime=0}^1  R_k \left( {v}_T, {{q}^{\,2} \over m_N^2},\left\{c_i^\tau c_j^{\tau^\prime} \right\} \right)
~W_k^{\tau \tau^\prime}( {q}^{\,2} b^2)
\label{eq:RS}
\end{equation}
where the sum extends over products of WIMP response functions $R_k$ and
nuclear response functions $W_k$.   The $R_k$ isolate the particle physics: they depend on
specific combinations of bilinears in the low-energy constants (LECs) of the EFT -- the 2$\cal{N}$ 
coefficients of Eq. (\ref{eq:Lag}) -- labeled by isospin $\tau$ (isoscalar, isovector).  The WIMP response functions also depend
on the magnitude of the WIMP velocity  ${v}_T$, measured relative to the center of mass of the
nucleus, and the three-momentum transfer 
$\vec{q}= \vec{p}^{\,\prime}-\vec{p}=\vec{k} - \vec{k}^{\prime}$.
The  nuclear response functions $W_k$ can be varied by experimentalists, if they explore a variety
of nuclear targets.   
The $W_k$ are functions of $y \equiv (q b/2)^2$, where $b$ is the nuclear size (explicitly the
harmonic oscillator parameter if the nuclear wave functions are expanded in that 
single-particle basis).  

\begin{figure}[t]
\includegraphics[width=160mm]{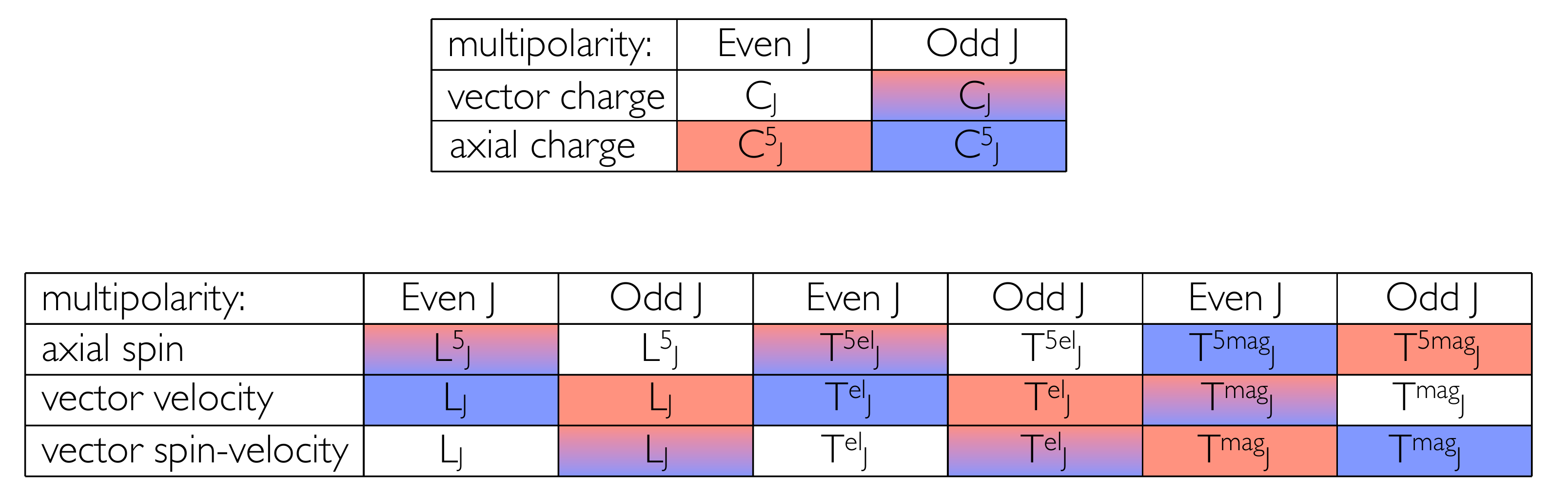}
\caption{Constraints imposed by the parity and time-reversal properties of the nuclear ground state on the multipole operators 
mediating WIMP-nucleus elastic scattering.  Transitions forbidden by parity (time reversal) are tinted red (blue); those forbidden
by both are indicated by the red-blue gradient fill.
Six types of multipole operators are allowed, corresponding to the even-angular-momentum-J multipoles of the vector charge operator (which generate the SI response),
the odd-J longitudinal and transverse electric multipoles of the axial spin operator (which generate in combination
the standard SD response), the odd-J transverse
magnetic multipoles of the vector velocity operator, and the even-J longitudinal and transverse electric multipoles of
the vector spin-velocity operator.}
\label{fig:fig1}
\end{figure}

The embedding of the WIMP-nucleon interaction in the nucleus will necessarily lead to the most
general form of the WIMP-nucleus interaction, provided the quantum mechanics is done properly.
But it is also satisfying to derive the end result from symmetry considerations alone.  Our EFT 
operators are constructed from the nucleon charge, spin, and velocity.  These operators can
be combined to form a nucleon vector charge operator $1(i)$, an axial-vector charge $\vec{\sigma}(i) \cdot \vec{v}(i)$,
an axial spin current $\vec{\sigma}(i)$, a vector velocity current $\vec{v}(i)$, and a vector
spin-velocity current $\vec{\sigma}(i) \times \vec{v}(i)$.  Charges and currents can be combined with
the plane-wave operator appearing in the scattering,
$ e^{i \vec{q} \cdot \vec{r}(i)} $,
to produce nuclear multipoles operators $\hat{O}_J$ carrying definite angular momentum and parity, and
transforming with a definite sign under time reversal.  In the case of the currents, one can take 
three projections -- longitudinal, transverse electric, and transverse magnetic -- relative to a
quantization axis defined by the momentum transfer.
As the nuclear ground state is nearly an eigenstate of both parity and time reversal, only those multipole
operators that are even under parity and time reversal can contribute.  By eliminating all operators with
the wrong transformation properties (a exercise familiar from standard weak interactions theory) one 
obtains the surviving response functions and thus the general form of the cross section.  The 
results, shown in Fig. \ref{fig:fig1}, indicate six response functions are needed to describe WIMP-nucleus
elastic scattering, not the two (SI/SD) usually assumed.

Specific forms for the WIMP and nuclear response functions appearing in Eq. (\ref{eq:RS}) can be obtained from the 
EFT Lagrangian, as detailed
in Ref. \cite{Nikhil2014}, under the assumption that the WIMP-nucleus interaction is the sum over the various EFT WIMP-nucleon
interactions.  As the arguments just given above indicate there must be six types of nuclear multipole operators
in a model-independent formulation, and as our effective theory Lagrangian gives the most general WIMP-nucleon
interaction, one might anticipate that a straight-forward application of that Lagrangian would then generate the explicit forms
of the nuclear multipole operators.  While this takes some algebra, indeed this is what happens.  The nuclear
response functions are generated by six operators that we can write, for simplicity, in their long-wavelength
forms (that is, without accompanying form factors).  One finds the leading-order multipoles
\begin{eqnarray}
C_{J=0M} \rightarrow \sum_{i=1}^A 1(i)~~~~~~~~~~~~~~~L_{J=1M}^5 \sim T_{J=1M}^{\mathrm{el}~5} \rightarrow \sum_{i=1}^A \sigma_{1M}(i)~~~~~~~~~~~~~~~T_{J=1M}^\mathrm{mag} \rightarrow {q \over m_N} \sum_{i=1}^A \ell_{1 M}(i)  ~~~~~~~~~~~~~~~~~~~~\nonumber \\
 L_{J=0M} \rightarrow {q \over m_N} \sum_{i=1}^A \vec{\sigma}(i) \cdot \vec{\ell}(i) ~~~~~~~~ L_{J=2M} \rightarrow  {q \over m_N} \sum_{i=1}^A \left[ r(i) \otimes \left( \vec{\sigma}(i) \times {1 \over i} \vec{\nabla} \right) \right]_{2M} ~~~~~~~~
  T^\mathrm{el}_{J=2M} \rightarrow  {q \over m_N} \sum_{i=1}^A \left[ r(i) \otimes \left( \vec{\sigma}(i) \times {1 \over i} \vec{\nabla} \right) \right]_{2M} \nonumber 
 \end{eqnarray} 
 The first two operators are the standard SI/SD interactions, though as indicated the effective
 theory divides the spin response into separate longitudinal ($L_{JM}^5$) and transverse ($T_{JM}^{\mathrm{el}~5}$) components.  These
 spin responses have distinct
 $R_k ( {v}_T^{ 2}, {q}^{\,2}/m_N^2 )$ -- different bilinear functions of the $c_i$ and thus different sensitivity to
 the underlying EFT operators -- as well as $W_k(q^2b^2)$ characterized by distinct form factors.
 In addition we find three new response functions, each proportional to $q/m_N$ (an
 indication that they are connected with the finite site of the nucleus), generated in two cases by
 familiar scalar and vector operators, $\vec{\sigma}(i) \cdot \vec{\ell}(i)$ and $\vec{\ell}(i)$, and in the third case by a
 tensor operator that is clearly closely related to $\vec{\sigma}(i) \cdot \vec{\ell}(i)$.  For one response function,
 the vector longitudinal response $L_{JM}$, both scalar and tensor terms survive in the long wavelength limit.
 
 One sees that the general form of the nuclear response, even in the long wavelength limit, is more
 complex than is apparent from the SI/SD formulation.  Two scalar ($J=0$) operators emerge from the effective theory,
 and three vector ($J=1$) operators.  Two sets of operators interfere, $C_{JM}-L_{JM}$ and $T_{JM}^{\mathrm{el}~5}-T_{JM}^\mathrm{mag}$, as happens in familiar electroweak processes such as neutrino scattering off nuclei.
 The bottom line for experimentalists is that each nuclear response function is a ``knob" that
 can be turned by picking nuclear targets with the requisite properties.  Each response function is accompanied
 by a WIMP tensor  $R_k ( {v}_T^{2}, {q}^{2}/m_N^2 )$ that involves a distinct set of LECs.  
 Thus, including the two interference terms, in principle one can obtain eight constraints on the LECs
 of the effective theory from elastic scattering experiments.  In the future, when dark matter
 scattering is observed, the experimental task will have been completed when those eight constraints
 are determined: all information on candidate fundamental theories (the ultra-violet theories) that can be extracted
 from inelastic scattering will have been obtained, at that point.  
 
 Experimentalists can turn the nuclear physics ``knobs" by selecting nuclear targets that isolate the various
 response functions, thereby determining the associated $R_k ( {v}_T^{2}, {q}^{2}/m_N^2 )$.  For example, if 
 one wants to extract the WIMP physics associated with the two response functions 
 generated by $\vec{\sigma}(i)$ and $\vec{\ell}(i)$, one could employ two odd-A nuclear targets,
 one having its unpaired valence nucleon in an $\ell=0$ orbit
 (an s-wave orbit), the other with a valence nucleon in a high-$\ell$ orbit.  The former would be sensitive only to
 $\vec{\sigma}(i)$, the latter primarily to $\vec{\ell}(i)$.  In fact, one should do four such experiments
 two with targets having unpaired valence protons and two with targets having unpaired valence neutrons, to test
 the isospin of these couplings.  With such a strategy, extended to the full set of nuclear response functions,
 one would eventually learn everything than can
 be learned from elastic scattering.
 
\section{The Point-Nucleus Limit: The Origin of the Additional Response Functions} 
 Before turning to the $R_k ( {v}_T^{2}, {q}^{2}/m_N^2 )$, it is interesting to explore the
 origin of the additional response functions.  The standard SI/SD form of the cross section can be obtained
 (apart from the detail that one should allow the longitudinal and transverse projections of spin to
 have distinct coefficients)
 by taking the point nucleus limit.  Naively one might hope that this point-nucleus SI/SD form corresponds
 to some lowest-order approximation, capturing the leading-order effects of candidate effective interactions.
 But in fact this is not the case for about half of the operators appearing in Eq. (\ref{eq:ops}), those involving the 
 WIMP-nucleon relative velocity.
 
 As it is easiest to illustrate the point in an example, consider
 \[  \mathcal{O}_8 = \sum_{i=1}^A \vec{S}_\chi(i) \cdot \vec{v}^\perp(i),~~~~~~~~~~\vec{v}^\perp(i) \equiv \vec{v}_\chi - \vec{v}_N(i). \]
(We have made a slight simplification by not symmetrizing the velocity between initial and final states -- 
see \cite{Liam2013} for details.)   If we take the point-nucleus limit, all nucleons are at the same point, and the
interaction reduces to 
\[  \vec{S}_\chi(i) \cdot \vec{v}_T \sum_{i=1}^A 1(i) ,\]
so that the nuclear interaction is SI with a coefficient that depends on the WIMP velocity relative to the
nuclear center-of-mass, $v_T \sim 10^{-3}$.  Associated cross sections, which depend on the 
square of this amplitude, would be of order $\sim 10^{-6}$ of the weak interaction value, consequently.
However it is easy to see that this point-nucleus limit is unjustified.  The original interaction
contains $A$ independent internal WIMP-nucleon velocities, which we are free to rewrite by a
standard Jacobi transformation as
\[ \{\vec{v}^\perp(i), i=1,....,A \}~ \rightarrow~ \{ \vec{v}_T;~ \vec{\dot{v}}(i),i=1,....A-1 \} \]
singling out the WIMP velocity relative to the nuclear center-of-mass, leaving
$A-1$ velocities $\vec{\dot{v}}(i)$ corresponding to the independent internucleon velocities.
For example, for the simplest case of $A$=2, there is one such 
internucleon velocity, $\vec{\dot{v}}_1 \equiv (\vec{v}(2)-\vec{v}(1))/\sqrt{2}$.
Now internucleon velocities are typically $\sim 10^{-1}$, not $\sim 10^{-3}$.  Thus the point-nucleus limit
effectively keeps one amplitude that is tiny (proportional to $v_T$), and discards $A-1$ other velocities that couple to $\vec{S}_\chi$
with precisely the same strength, that are 100 times larger.

One might be mislead into thinking that one can neglect the internucleon velocities, despite their size,
because of parity:  they are intrinsic nuclear operators carrying odd parity.  But this overlooks the fact that
while the energy transfer in WIMP scattering is small, the three-momentum transfer is large, typically
$q r(i) \gtrsim 1$.  Thus expanding the full nuclear operator to first order in q (while quantizing along
$\vec{q}$ and taking the $J=1$ part) yields
\[ e^{i \vec{q} \cdot \vec{r}(i)} \vec{\dot{v}}(i)~ \rightarrow~ -{1 \over i} q \vec{r}(i) \times \vec{\dot{v}}(i) =
-{1 \over i} {q \over m_N} \vec{r}(i) \times \vec{\dot{p}}(i) = -{q \over m_N} \vec{\ell}(i) \]
One obtains a familiar dimensionless parity-conserving operator, the orbital angular momentum,
accompanied by a dimensionless constant $q/m_N \sim 1/10 $.  That is, our derivation shows that
the scale of the internal relative nucleon velocities is encoded in the parameter $q/m_N$.

This is then very helpful in the effective theory formulation.  We mentioned previously that the most general Hermitian
WIMP-nucleon interaction can be constructed from the four dimensionless variables
\[ \vec{S}_\chi~~~~~\vec{S}_N~~~~~\vec{v}^\perp~~~~~i{\vec{q} \over m_N} \]
The reason that $m_N$ is the natural choice for the mass scale in the fourth variable is that the 
parameter $q/m_N$ naturally arises from the proper treatment of relative nucleon velocities in the nucleus.
As outlined below (and discussed in greater detail in \cite{Nikhil2014}), when velocity-dependent
interactions are treated in a manner that properly takes into account nuclear
finite size, their effects are enhanced by a factor 
of $\sim (\mu_T/m_N)^2 \sim 10^4$, where $\mu_T$ is the reduced mass for the WIMP-nucleus
scattering, relative to treatments that take the point-nucleus limit.  
Thus velocity-dependent interactions that appear to be very difficult to test, in the point nucleus
limit, in fact make large contributions.  These enhancement are connected with
new response functions generated by operators like $\vec{\ell}(i)$: by exploiting these new response functions,
one can determine the strengths of such velocity-dependent interactions.  In addition to misrepresenting the
 magnitude of the cross section,
the point nucleus limit also mischaracterizes the multipolarity of the scattering.  This is seen in
our example, $\vec{S}_\chi \cdot \vec{v}^\perp$.  The associated point-nucleus operator is $1(i)$,
a scalar, while the scattering is actually dominantly vector, generated by $\vec{\ell}(i)$.

\section{Constraining the Effective Theory Parameters} 
The WIMP-nucleus differential cross section as a function of the recoil energy $E_R$ carried off by the
scattered nucleus can be written
\begin{equation}
{ d \sigma(v_T,E_R) \over dE_R} = {2 m_T \over \pi v_T^2} G_F^2 \left[ {1 \over 2 j_\chi+1}{1 \over 2j_N+1} \sum_\mathrm{spins} \left| \mathcal{M}^\mathrm{Nuc} \right|^2 \right]=  {2 m_T \over \pi v_T^2} G_F^2 {4 \pi \over 2j_N+1}
\sum_{\tau,\tau^\prime=0}^1 \sum_k R_k^{\tau \tau^\prime} \left( v_T^{ 2}, {q^{\,2} \over m_N^2}  \right)
~W^{\tau \tau^\prime}_k( q^{\,2} b^2)
\end{equation}
where the sum extends over the various response functions, generated by the six sets of multipole operators
including two interference terms.  The Fermi constant has been introduced to make the weak scale explicit.  The
sums over the isospin indices $\tau,\tau^\prime$ appear because the 
$R_k^{\tau \tau^\prime}$ depend on the LECs
\[ c_i^{\tau=0} =(c_i^{(p)} + c_i^{(n)})/2~~~~~~~~~~  c_i^{\tau=1} =(c_i^{(p)} - c_i^{(n)})/2  \]
we introduced previously.  Because we have normalized these to the weak scale,
the LECs are dimensionless.  The conventions employed here
are those of Ref. \cite{Nikhil2014}, where more details are given.

In this formula, the left hand side will be measured (some day, hopefully), while on the right hand side, the experimentalist
can vary the $W^{\tau \tau^\prime}_k( q^{\,2} b^2)$, the nuclear response functions.  These 
response functions, as $q \rightarrow 0$, are generated by squaring the nuclear matrix elements 
of the long-wavelength operators discussed in the
previous section; for general $q$, they
also involve nuclear form factors,
which in the case of the harmonic oscillator shell model, depend on $(qb)^2$, where $b$ is the harmonic oscillator size parameter.  Consequently, in an ideal world, experimentalists would vary the $W^{\tau \tau^\prime}_k( q^{\,2} b^2)$
by using nuclear targets with different angular momenta J, different isospins, and different shell structures, to obtain
a sufficient number of constraints 
to map out the eight $R_k^{\tau \tau^\prime} \left( v_T^{ 2}, q^{\,2}/m_N^2  \right)$ (six response functions
plus two interference terms).  One would have then
obtained all of the information on dark matter particle interactions that is available from elastic scattering.
Note that more than eight constraints can be obtained:  the measurements can be done as a function of $E_R =q^2/2 m_T$,
$m_T$ the target mass, to separate the LECs contributing to a given $R_k^{\tau \tau^\prime} \left( v_T^{ 2}, {q^{\,2} \over m_N^2}  \right)$.

It is helpful to consider examples in which we contrast the standard SI/SD treatment with a full quantum mechanical
treatment, while still employing the EFT Lagrangian.  For SI scattering we find
\begin{equation}
R_{SI}^{\tau \tau^\prime} = c_1^\tau c_1^{\tau^\prime} +{j_\chi(j_\chi+1) \over 3} \left[ {q^2 \over m_N^2} v_T^2 c_5^\tau c_5^{\tau^\prime}+v_T^2 c_8^\tau c_8^{\tau^\prime} + {q^2 \over m_N^2} c_{11}^\tau c_{11}^{\tau^\prime} \right]
~~~~~W^{\tau \tau^\prime}_{SI}  \underset{q \rightarrow 0} {\longrightarrow} {1 \over 4 \pi} \langle j_N || \sum_{i=1}^A t^\tau(i) ||j_N \rangle \langle j_N ||\sum_{i=1}^A t^{\tau^\prime}(i) || j_N \rangle
\end{equation}
where $j_\chi$ is the WIMP spin.  We see that the EFT operators  $\mathcal{O}_5$ and $\mathcal{O}_8$ 
contribute to SI scattering, but their contributions are suppressed by $\sim 10^{-8}$ and $\sim 10^{-6}$ respectively.
($\mathcal{O}_8$ was the velocity-dependent operastor example we discussed in the previous section.)  But if
the nuclear finite size is properly treated, one obtains a $T^\mathrm{mag}$ response function as well
\begin{equation}
R_{T^\mathrm{mag}}^{\tau \tau^\prime} = {j_\chi (j_\chi+1) \over 3}\left[ {q^2 \over m_N^2} c_5^\tau c_5^{\tau^\prime}+c_8^\tau c_8^{\tau^\prime} \right]~~~~~W^{\tau \tau^\prime}_{T^\mathrm{mag}}  \underset{q \rightarrow 0} {\longrightarrow} {1 \over 24 \pi}
{q^2 \over m_N^2} \langle j_N || \sum_{i=1}^A  \ell(i) ~t^\tau(i) ||j_N \rangle \langle j_N ||\sum_{i=1}^A \ell(i) ~ t^{\tau^\prime}(i) || j_N \rangle
\end{equation}
The two EFT operators $\mathcal{O}_5$ and $\mathcal{O}_8$ now are associated with kinematic suppression 
factors of $10^{-4}$ and $10^{-2}$, respectively.  Comparing the two results, we see that a proper treatment of
finite size
\begin{itemize}
\item  allows us to separately measure the effects of $c_5$ and $c_8$, when previously
they appeared in combination with $c_1$ and $c_{11}$; 
\item generates a response with increased sensitivity to
these two couplings; and 
\item leads to a dominant rank-one response requiring $j_N>0$, in contrast to the point-nucleus result that
$\mathcal{O}_5$ and $\mathcal{O}_8$ contribute very weakly to the scalar SI response.
\end{itemize}

Similar in the SD point-nucleus limit
\begin{eqnarray} 
R_{SD}^{\tau \tau^\prime}={1 \over 12} \left[ {q^2 \over m_N^2} v_T^2 c_3^\tau c_3^{\tau^\prime}+v_T^2 c_7^\tau c_7^{\tau^\prime}+{q^2 \over m_N^2} c_{10}^\tau c_{10}^{\tau^\prime} +{j_\chi(j_\chi+1) \over 3} \left( 3 c_4^\tau c_4^{\tau^\prime} +{q^2 \over m_N^2} (c_4^\tau c_6^{\tau^\prime}+c_6^\tau c_4^{\tau^\prime}  
 +2 c_9^\tau c_9^{\tau^\prime})  \right. \right. \nonumber \\  \left. \left.  +{q^4 \over m_N^4} c_6^\tau c_6^{\tau^\prime}+2 v_T^2 c_{12}^\tau c_{12}^{\tau^\prime} +{q^2 \over m_N^2} v_T^2 (c_{13}^\tau c_{13}^{\tau^\prime} + c_{14}^\tau c_{14}^{\tau^\prime}-c_{12}^\tau c_{15}^{\tau^\prime} -c_{15}^\tau c_{12}^{\tau^\prime} ) + {q^4 \over m_N^4} v_T^2 c_{15}^\tau c_{15}^{\tau^\prime} \right) \right] \nonumber \\
 W^{\tau \tau^\prime}_{SD}  \underset{q \rightarrow 0} {\longrightarrow} {1 \over 4 \pi}
 \langle j_N || \sum_{i=1}^A  \sigma(i) ~t^\tau(i) ||j_N \rangle \langle j_N ||\sum_{i=1}^A \sigma(i) ~ t^{\tau^\prime}(i) || j_N \rangle~~~~~~~~~~~~~~~~~~~~~~~~~~~~~~~~~~~~~~~~~~~~~~~~~~~~~~~~~~~~~~
\end{eqnarray}
But the velocity-suppressed operators appearing above make much stronger contributions to, and can be extracted
more cleanly from, the new response functions that arise from properly treating the structure of the nucleus.
For example, for scattering off a $j_N=0$ nucleus, the longitudinal response function is 
\begin{eqnarray}
R_L={q^2 \over4  m_N^2}c_3^\tau c_3^{\tau^\prime} + {j_\chi(j_\chi+1) \over 12} \left(c_{12}^\tau -{q^2 \over m_N^2} c_{15}^\tau \right) \left(
c_{12}^{\tau^\prime} - {q^2 \over m_N^2} c_{15}^{\tau^\prime} \right)~~~~~~~~~~~~~~~~~~~~ \nonumber \\
W^{\tau \tau^\prime}_{L}  \underset{q \rightarrow 0} {\longrightarrow} {1 \over 36 \pi}
{q^2 \over m_N^2} \langle j_N || \sum_{i=1}^A  \vec{\sigma}(i) \cdot \vec{ \ell}(i) ~t^\tau(i) ||j_N \rangle \langle j_N ||\sum_{i=1}^A \vec{\sigma}(i) \cdot \vec{\ell}(i) ~ t^{\tau^\prime}(i) || j_N \rangle
\end{eqnarray}
Here, as in our previous example, $\mathcal{O}_3$, $\mathcal{O}_{12}$, and $\mathcal{O}_{15}$ make substantially larger
contributions to a new scalar response than they do to the SD response.  In a SI/SD treatment, one would 
conclude that these operators appear in a complicated way
in the SD response, entwined with $\mathcal{O}_4$, $\mathcal{O}_6$, $\mathcal{O}_7$, $\mathcal{O}_{9}$, $\mathcal{O}_{10}$, $\mathcal{O}_{13}$, and $\mathcal{O}_{14}$ and suppressed by $v_T^2$,
when in fact they make the leading
contributions to a new type of scalar response.  In principle all three couplings could be determined from the
nuclear recoil distribution, as 
$\mathcal{O}_3$, $\mathcal{O}_{12}$, and $\mathcal{O}_{15}$ appear in $R_L$ with distinct leading-order behaviors in
$q^2/m_N^2$.

These few examples show that many more tools become available for constraining the effective theory's LECs once the
new response functions associated with nuclear compositeness are recognized.  Velocity-dependent interactions
that appear to be difficult to probe in the standard SI/SD formalism can be isolated and studied in these
new response functions, where they play enhanced roles.

\section{Implications for Experiments} 
In the previous section we have noted that the larger number of responses that are revealed when the
nuclear finite size is properly treated opens up lots of additional opportunities for constraining the
parameters of the effective theory, and thus for restricting the form of candidate ultraviolet theories
that map onto the EFT.  Conversely, if one limits consideration to the standard SI/SD description
of the scattering, one can be mislead.  For example, conflicts may appear to arise
between experiments because the analysis framework is too restrictive, when in fact no conflict exists.

\begin{figure}[t]
\includegraphics[width=150mm]{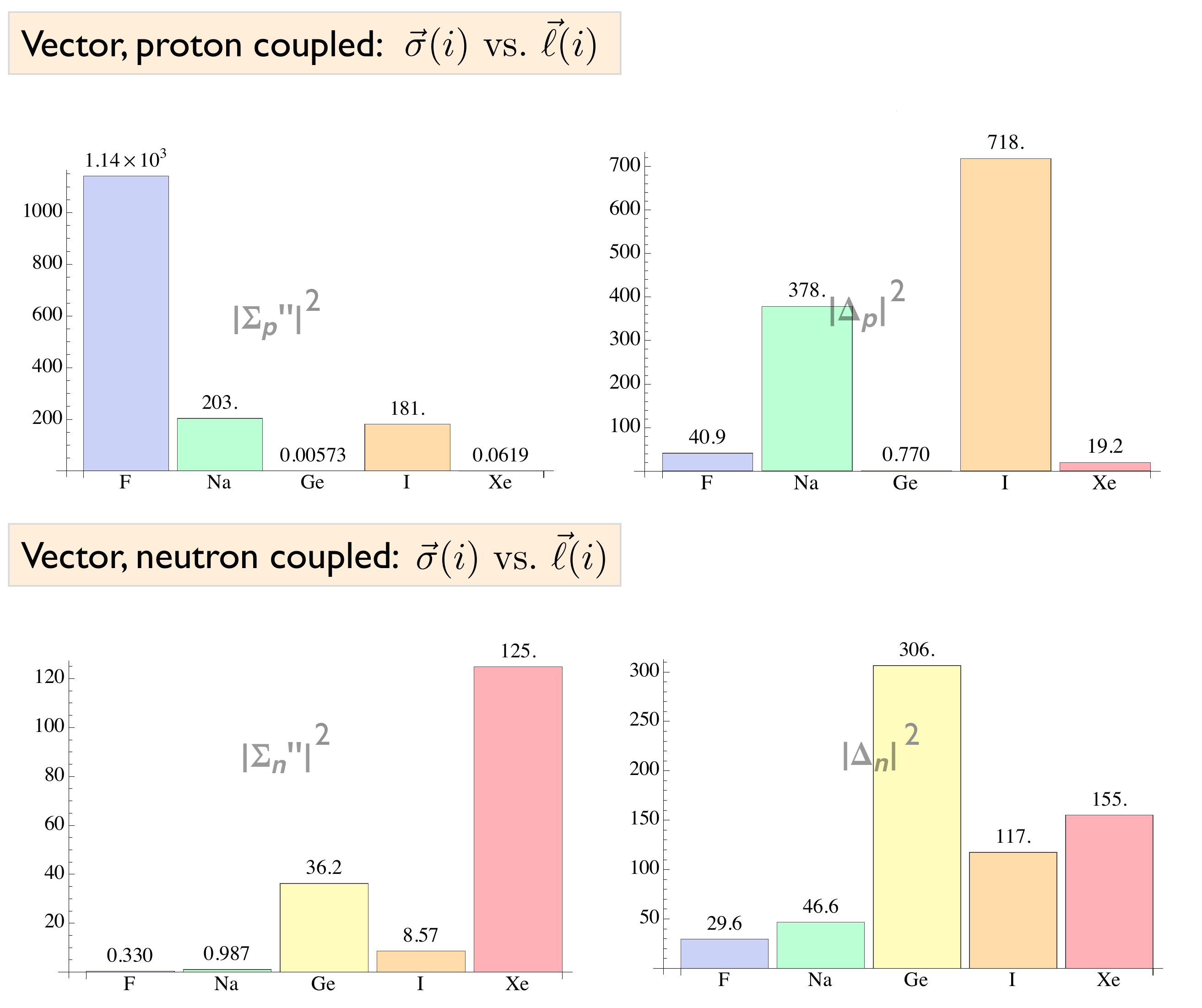}
\caption{A comparison of the spin response function (the longitudinal and transverse electric components of an
axial spin operator, left frames), employed in SI/SD analyses, with the orbital angular momentum response function
(the transverse magnetic component of the vector velocity operator, right frames).  The upper (lower) frames assume a
coupling to protons (neutrons).  The calculations are taken from the shell-model work of \cite{Liam2013,Nikhil2014}. }
\label{fig:fig2}
\end{figure}

One illustration in given in Fig. \ref{fig:fig2}, where the spin response function used in standard SI/SD 
responses is compared to the transverse magnetic (orbital angular momentum)  response function, for the targets F, Na, Ge, I, and Xe.  The results
come from the shell-model work of \cite{Liam2013}.  Both response functions are ``SD" in the sense that they
involve rank-one operators -- they cannot be distinguished geometrically -- but as they depend on distinct
underlying nuclear operators, they can be distinguished by selecting targets with the right properties.
In fact there are large differences in the responses to the
two underlying operators for targets currently in use in dark matter studies.  For example, if the operators couple to protons,
the ratio in the responses of iodine (e.g., DAMA, KIMS, COUPP) 
 and fluorine (e.g., PICASSO, PICO, COUPP) for $\langle \vec{\sigma} \rangle^2$ vs. $\langle \vec{\ell} \rangle^2$
differ by $\sim$ 110.  For couplings to neutrons, the responses of Xe (e.g., LUX, Xenon, XMASS) are similar for spin and orbital
angular momentum, but those for I and Ge (e.g., CDMS, COGENT, Edelweiss)  change by an order of magnitude, while those for
Na (e.g., DAMA, ANAIS) and F change by two orders of magnitude.   Consequently, if one analyzes experiments assuming
a SD coupling, but in fact the underlying operator is $\vec{\ell}(i)$, experiments that are
consistent may appear to be in conflict.   Properly interpreted, however, one would find
instead evidence that the underlying WIMP-nucleon interaction is
associated with $\vec{\ell}(i)$, not $\vec{\sigma}(i)$: this ansatz could then be tested in a follow-up experiment.

The field is entering the ``G2" phase of dark matter elastic scattering experiments: a smaller number
of experiments may be done, but with larger detector masses.  Our effective theory arguments, however,
argue for continued development of detectors using a variety of nuclear targets.  If dark matter is seen
in direct-detection experiments, there is no doubt that a large set of experiments will be needed:
as established here, a great deal of information can be extracted from such experiments because of
the multiple response functions that contribute.  Thus it is important, even if only a few detectors
are in operation, to continue developing new technologies that can be deployed when needed.
One technique and one target will not answer all of the question.  Even early
in the discovery phase, it will be important to analyze experiments in the kind of general framework 
provided here.  Repeating an example from the paragraph above, if rates in a Ge detector appear to be high compared to those seen
in a Xe detector, this might suggest $\vec{\ell}(i)$ as an operator and thus influence the choice of a 
third detector.

We would like to acknowledge useful conversations with Spencer Chang, Tim Cohen, Eugenio Del Nobile, Paddy Fox, Ami Katz, and Tim Tait. NA thanks the UC MultiCampus Research Program Dark Matter Search Initiative
for support.  ALF was partially supported by ERC grant BSMOXFORD no. 228169. WH is supported by the US Department of Energy under contracts DE-SC00046548 and DE-AC02-98CH10886.






\begin{thebibliography}{00}
\bibitem{Freese2013} K. Freese, M. Lisanti, and C. Savage, {\it Rev. Mod. Phys.} {\bf 85}, 1561 (2013).

\bibitem{Bertone2005} G. Bertone, D. Hooper, and J. Silk, {\it Phys. Repts.} {\bf 405}, 279 (2005).

\bibitem{Goodman1985} M. W. Goodman and E. Witten, {\it Phys. Rev. D} {\bf 31}, 3059 (1985).

\bibitem{Drukier1986}  A. K. Drukier, K. Freese, and D. N. Spergel, {\it Phys. Rev. D} {\bf 33}, 3495 (1986).

\bibitem{Jungman1996} G. Jungman, M. Kamionkowski, and K. Griest, {\it Phys. Repts.} {\bf 267}, 195 (1996).

\bibitem{Smith1990}  P. Smith and J. Lewin, {\it Phys. Rept.} {\bf 187}, 203 (1990).

\bibitem{Lewin1996} J. Lewin and P. Smith, {\it Astropart. Phys.} {\bf 6}, 87 (1996).

\bibitem{Fan2010} J. Fan, M. Reece, and L.-T. Wang, {\it JCAP} {\bf 11}, 042 (2010).

\bibitem{Liam2013} A. Liam Fitzpatrick, W, C. Haxton, E. Katz, N. Lubbers, and Y. Xu, {\it JCAP} {\bf 004}, 1302 (2013).

\bibitem{Hill2013} R. J. Hill and M. P. Solon, arXiv:1309.4092.

\bibitem{Nikhil2014} N. Anand, A. Liam  Fitzpatrick, and W. C. Haxton, to be published in {\it Phys. Rev. C}
(arXiv: 1308.6288).



\end{thebibliography}







\end{document}